\documentclass[aip,apl,amsmath,amssymb,linenumbers,reprint]{revtex4-1}
\usepackage{graphicx}
\usepackage[version=3]{mhchem} 
\usepackage{subfig}
\usepackage{dcolumn}
\usepackage{bm}

\begin{document}

\title{Structural and Electronic Properties of Hybrid Perovskites for High-Efficiency Thin-Film Photovoltaics from First-Principles}

\author{Federico Brivio}
\affiliation{Centre for Sustainable Chemical Technologies and Department of Chemistry, University of Bath, Claverton Down, Bath BA2 7AY, UK}

\author{Alison B. Walker}
\affiliation{Department of Physics, University of Bath, Claverton Down, Bath BA2 7AY, UK}

\author{Aron Walsh}
\email[Electronic mail:]{a.walsh@bath.ac.uk}
\affiliation{Centre for Sustainable Chemical Technologies and Department of Chemistry, University of Bath, Claverton Down, Bath BA2 7AY, UK}

\date{\today}

\begin{abstract}
The performance of perovskite solar cells recently exceeded 15 \% solar-to-electricity conversion efficiency for small-area devices. The fundamental properties of the active absorber layers, hybrid organic-inorganic perovskites formed from mixing metal and organic halides [\textit{e.g.} (NH$_4$)PbI$_3$ and (CH$_3$NH$_3$)PbI$_3$], are largely unknown. The materials are semiconductors with direct band gaps at the boundary of the first Brillouin zone. The calculated dielectric response and band gaps show an orientation dependence, with a low barrier for rotation of the organic cations. Due to the electric dipole of the methylammonium cation, a photoferroic effect may be accessible, which could enhance carrier collection.  
\end{abstract}


\maketitle 


Progress in the performance of hybrid pervoskite solar cells has rapidly advanced over the last five years.\cite{kojima-6050,snaith-1,gratzel-1,gratzel-2,heo-486,snaith-2} They represent the convergence of inorganic thin-film and dye-sensitised solar cells.\cite{park-2423} The conversion efficiencies have quickly surpassed both those of conventional dye-cells, as well as next-generation thin-film absorbers such as Cu$_2$ZnSnS$_4$.\cite{todorov-156,walsh-400} Despite their high-performance for small area (\textit{ca.} 0.2 cm$^2$) cells, the underlying materials properties are largely unknown, which could help with producing more robust large-area devices.

\textit{Perovskite} refers to the crystal structure of the mineral CaTiO$_3$, which is adopted by a large family of ABX$_3$ materials, with two notable examples being SrTiO$_3$ and BaTiO$_3$ (Figure 1). They are well known for their phase complexity, with accessible cubic, tetragonal, orthorhombic, trigonal and monoclinic polymorphs depending on the tilting and rotation of the BX$_6$ polyhedra in the lattice.\cite{glazer-1972} Phase transitions are frequently observed under the influence of temperature, pressure and/or electric field. 

While oxide perovskites (ABO$_3$) are formed from divalent A$^{II}$ ($1b$ site -- cuboctahedral) and tetravalent B$^{IV}$  ($1a$ site -- octahedral)  metals, halide perovskites (\textit{e.g.} ABI$_3$) can be formed from monovalent A$^{I}$ and divalent B$^{II}$ metals. For example, the application of CsSnI$_3$ in solar cells has recently been demonstrated.\cite{chung-486} Hybrid organic-inorganic perovskites are produced by replacing one of the inorganic cations by an isovalent molecule, \textit{e.g.} CsSnI$_3$ $\rightarrow$ (NH$_4$)SnI$_3$, where NH$_4^+$ (A) is the ammonium cation. Recently, the methylammonium   (MA) cation (\textit{i.e.} CH$_3$NH$_3^+$) has been widely applied, resulting in the highest-performance perovskite-structured photovoltaic absorbers.\cite{snaith-1,gratzel-1} More generally, a large series of hybrid perovskites have been reported, which vary in their dimensionality and the orientation of the underlying perovskite lattice.\cite{calabrese-2328,mitzi-1,mitzi-2,borriello-235214}  

\begin{figure}[hb]
\begin{center}
\resizebox{8.5cm}{!}{\includegraphics*{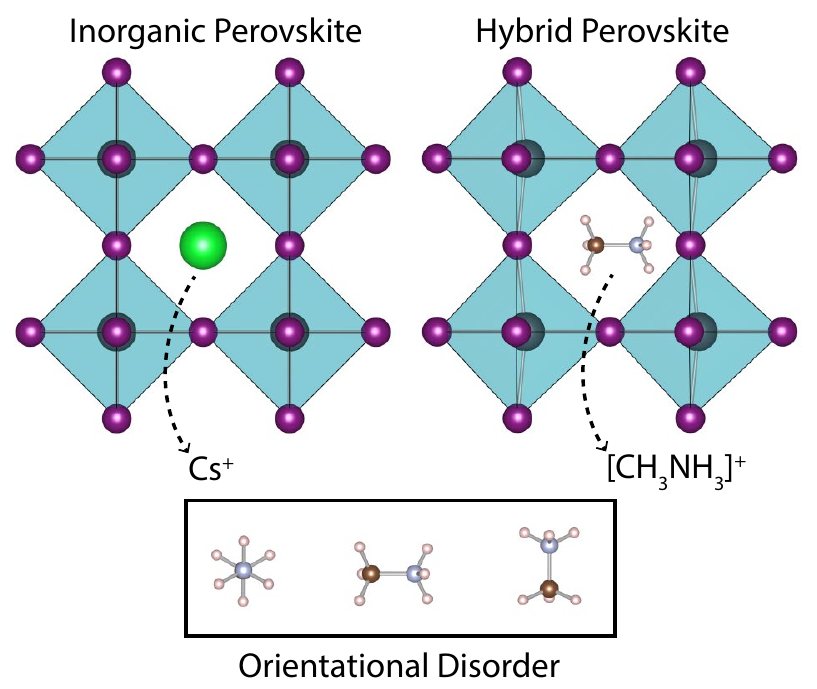}}
\caption{\label{fig1} Illustration of the perovskite structure based on corner sharing octahedra of BX$_6$ with either a monovalent metal (inorganic) or charged molecule (hybrid) at the centre of the unit cell. For hybrids, there is an orientation dependence on the central cation.} 
\end{center} 7
\end{figure}

In this Letter we assess the properties of the two archetypal hybrid perovskites (NH$_4$)PbI$_3$ and (CH$_3$NH$_3$)PbI$_3$ using density functional theory (DFT) for the ground-state properties and density functional perturbation theory (DFPT) for the dielectric and optical response functions. We provide insights into the key properties required for device models and screening procedures and that underpin their utility in photovoltaic cells.

Characterisation of the crystal structures of inorganic perovskites is difficult, and hybrid perovskites are even more challenging. Recent analysis of high-quality (MA)PbI$_3$ crystals identified cubic ($Pm\bar{3}m$), tetragonal ($I4/mcm$) and orthorhombic ($Pnam$) phases from X-ray diffraction, while transmission electron microscopy suggested a pseudo-cubic behaviour that is consistent with variability in the octahedral titling and/or rotations.\cite{baikie-5628} For our investigation, we take a cubic basis, starting from the $Pm\bar{3}m$ lattice, and investigate the potential energy landscape associated with the molecular orientation. 

The crystal structures (internal forces and external pressure) were optimised at the level of DFT, with the exchange-correlation functional of Perdew, Burke and Ernzerhof revised for solids (PBEsol).\cite{pbesol}  The Pb $5d$ orbitals were treated as valence and scalar-relativistic effects are included. Further calculations were made using a non-local hybrid exchange-correlation functional (HSE06).\cite{hse06} Calculations were performed using the VASP code,\cite{vasp1} a 500 eV plane-wave cut-off, and reciprocal space sampling of $6\times6\times6$ $k$-point density. 
Internal structural parameters were converged to within 5 meV/\AA, and the phonon frequencies were checked at the zone-centre to ensure that no imaginary modes were present. 
The high-frequency ($\epsilon_{\infty}$) and static ($\epsilon_0$) dielectric constants were computed using DFPT\cite{dfpt} based upon a tightly converged electronic wavefunction (within $10^{-9}$ eV) and a denser grid of $10\times10\times10$ $k$-points.

For (A)PbI$_3$ there is no strong orientation dependence of the ammonium ion owing  to its approximately spherical topology. A local minimum is found where the four hydrogens are directed towards interstitial positions. For (MA)PbI$_3$ the behaviour is more subtle owing to a molecular dipole (of strength 2 D) associated with the methylammonium ion (\textit{i.e.} $[CH_3]^{\delta+}-[NH_3]^{\delta-}$). We identified three local minima, where the dipole is aligned along of the $<100>$, $<110>$ and $<111>$ directions, relative to the origin of the cubic lattice (Figure 1). The total energy difference is within 15 meV per atom, with the $<100>$ orientation being most stable. The absence of a significant barrier to rotation ($<$40 meV) is consistent with labile movement under standard conditions. Indeed, $^2$H and $^{14}$N spectra have confirmed that MA cation rotation is a rapid process at room temperature.\cite{wasylishen-58}

The predicted lattice constants for (MA)PbI$_3$ vary from 6.26 - 6.29 \AA  ~ (Table \ref{tbl:properties}) depending on the molecular orientation. These compare well to the value of 6.26 \AA ~ obtained from powder diffraction measurements\cite{baikie-5628} and 6.33 \AA ~ from computations.\cite{mosconi-13902} Distortions of the PbI$_6$ octahedra are observed for both materials, even within the cubic basis. The presence of the molecule lowers the internal lattice symmetry of the parent perovskite structure and allows octahedral tilting to take place. The inherent structural  `softness' of Pb(II) is well documented, arising from the stereochemical activity of the $6s^2$ lone pair electrons.\cite{walsh-4455}

\begin{figure}[hb]
\begin{center}
\resizebox{8.5cm}{!}{\includegraphics*{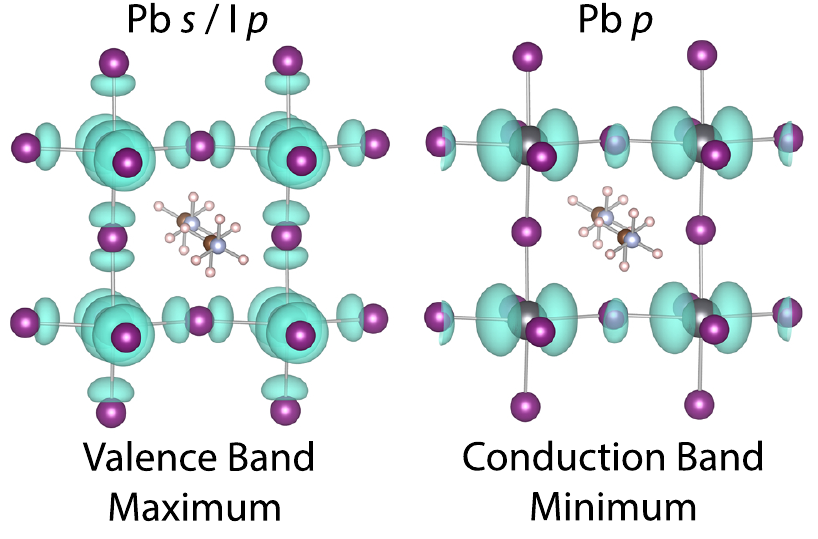}}
\caption{\label{fig2} Isosurface plot of the self-consistent electron density associated with the PBEsol wave functions of the upper valence and lower conduction bands of (MA)PbI$_3$. } 
\end{center}
\end{figure}

The electronic band gap is determined by the states at the valence band maximum (VBM) and conduction band minimum (CBM). For both materials, the VBM is formed of an anti-bonding Pb $s$/I $p$ combination, while the CBM is formed of empty Pb $p$ orbitals, consistent with the formal electronic configuration of $5d^{10}6s^26p^0$. From the topology of the band-edge wave functions (Figure 2) isotropic hole and electron band conduction is expected. The strong dispersion of the bands in reciprocal space, confirmed from E(\textit{k}) plots with $m^*_n$ and $m^*_p$ $<$ 0.3 $m_e$, is consistent with the reports of effective bipolar electrical conductivity.\cite{stoumpos-9019} A dedicated study of the defect physics is required to fully assess the origin of intrinsic conductivity. 

\begin{table*}[ht]
\small
  \caption{Predicted materials properties of hybrid perovskites, from density functional theory (PBEsol functional), where (MA) refers to the CH$_3$NH$_3$ cation. The diagonal of the dielectric tensors are given as $\epsilon^{xx}$,$\epsilon^{yy}$ and $\epsilon^{zz}$.}
  \label{tbl:properties}
\begin{ruledtabular}
\begin{tabular}{lcccc}
    \hline
Material & $a$ (\AA) & $E_g$ (eV) & $\epsilon_0$ & $\epsilon_{\infty}$ \\
    \hline
(NH$_4$)PbI$_3$        & 6.21  & 1.20 & 18.62, 18.47, 18.14 & 6.49, 6.49, 6.47 \\
(MA)PbI$_3$ -- $<100>$ & 6.29  & 1.38 & 22.39, 27.65, 17.97 & 6.29, 5.89, 5.75 \\
(MA)PbI$_3$ -- $<110>$ & 6.26  & 1.37 & 17.95, 23.56, 22.67  &  5.62, 6.54, 6.26 \\
(MA)PbI$_3$ -- $<111>$ & 6.28  & 1.37 & 36.52, 37.28, 24.94 &  6.10, 6.10, 5.92 \\
    \hline
\end{tabular}
\end{ruledtabular}
\end{table*}

The  magnitude of the band gap determines the onset of optical absorption and is closely related to the maximum voltage achievable in a photovoltaic device. The gap increases from 1.20 eV for (A)PbI$_3$ to 1.38 eV for (MA)PbI$_3$ (Table \ref{tbl:properties}).  In all cases, the band gap is strongly direct and determined at the boundary of the first Brillouin zone ($R:\frac{2\pi}{a}[\frac{1}{2},\frac{1}{2},\frac{1}{2}]$). The increase is related to the significant expansion of the cell due to the larger cation: there is no evidence of M or MA contributing electronically to the frontier orbitals. 
In agreement with our results, a previous investigation of Sn-based hybrids showed that while the organic cations maintain overall charge neutrality, implying electronic mixing with the framework, they do not contribute the upper valence or lower conduction bands states responsible for conductivity.\cite{borriello-235214} 
Indeed, the $\sigma$/$\sigma*$ bonds of A and MA are found at least 5 eV below the highest occupied state and 2.5 eV  above the lowest empty state, respectively. 

Although the application of hybrid perovskites in solar cells has predominately focused on (MA)PbI$_3$, our results suggest that (A)PbI$_3$ would also be effective for sensitisation towards longer wavelengths, and an (A)$_{x}$(MA)$_{1-x}$PbI$_3$ alloy could be used to tune the absorption onset. 
It should be noted that the same trends calculated with PBEsol are observed using the hybrid HSE06 functional; however, the predicted band gaps are larger. Optical absorption measurements place the band gap of (MA)PbI$_3$ at 1.5 eV\cite{baikie-5628}, while PBEsol and HSE06 predict values close to 1.4 eV and 2.0 eV, respectively. The calculated band gaps for materials formed by $ns^2$ cations such as Pb(II) are reasonably described without non-local electron exchange.\cite{wei-13605,walsh-1284} This originates from a cancellation of errors with the neglect of spin-orbit splitting in the I $5p$ valence and Pb $6p$ conduction bands.\cite{perovskites} 

Due to the pseudo-cubic lattice symmetry, the dielectric response of the hybrid materials is anisotropic (Table \ref{tbl:properties}). The high-frequency optical constants (5.6 -- 6.5) are close to those of other absorber materials, \textit{e.g.} $\epsilon_{\infty}^{CdTe} = 7.1$;\cite{madelung-04} however, the static dielectric response is much larger. The large dielectric constants of pervoskite materials are associated with the structural flexibility that can support ferro-, anti- and para-electric order. 
(A)PbI$_3$ is predicted to have low-frequency constants from 18.1 -- 18.6, while (MA)PbI$_3$ exhibits a stronger screening of 18.0 -- 37.3, which can be understood through the additional response from the molecular dipole in (MA)PbI$_3$. An isotropic average of the tensor across each orientation results in an effective dielectric constant of 25.7 for (MA)PbI$_3$, which is significantly larger than most tetrahedral semiconductors (\textit{e.g.} $\epsilon_{0}^{CdTe} = 10.4$)\cite{madelung-04} and will affect electron transport in a photovoltaic device, with stronger screening of any macroscopic electric fields across the absorber layer.

An order-disorder transition associated with the molecular dipole of MA, in a perovskite analogue, has recently been reported to give rise to reversible switching in the dielectric response.\cite{zhang-5230} While no evidence has been reported of this effect in hybrid halide perovskites, it is expected that they will also display some degree of amphidynamic behaviour arising from the collective motion of the constituent polar molecules. 

In summary, we have reported key properties of the hybrid perovskite layers used in thin-film solar cells. These materials combine low-energy direct band gaps with static dielectrics constants larger than 18. Due to the low barrier for rotation of the organic cations, no long-range ordering of the molecular dipoles associated with CH$_3$NH$_3$ is expected; however, this could change in the presence of an external electric field. A transition from para- to ferroelectric order would enhance electron-hole separation. If the electric field was light induced, \textit{e.g.} in a photovoltaic device, this could give rise to a novel photoferroic effect. 

We thank L.M. Peter, K.T. Butler and P.J. Cameron for useful discussions, and acknowledge membership of the UK's HPC Materials Chemistry Consortium, which is funded by EPSRC grant EP/F067496. F.B. is funded through the EU DESTINY Network (Grant 316494). A.B.W. has received funding from EPSRC grant EP/J017361/1, and A.W. acknowledges support from the Royal Society and the ERC (Grant 277757).


\begin{thebibliography}{29}
\bibitem [{per()}]{perovskites}%
  \BibitemOpen
  \href@noop {} {}\bibinfo {note} {A more accurate treatment than PBEsol would
  be HSE06+SOC; however, this was not possible due to the high computational
  cost.}\BibitemShut {Stop}%
\end{thebibliography}

%




\end{document}